\documentclass{article}    

\usepackage{graphicx}

\title{\textbf{The Cat nRules}}  
\author{Richard Mould\footnote{Department of Physics and Astronomy, State University of New York, Stony Brook,
\mbox{New York} 11794-3800; http://ms.cc.sunysb.edu/\~{}rmould}}  
\date{}    
   
\begin{document}             

\maketitle              

\begin{abstract}

     The nRules that are developed in another paper are applied to two versions of the Schr\"{o}dinger cat
experiment.  In version I the initially conscious cat is made unconscious by a mechanism that is initiated by
a radioactive  decay.  In version II the initially unconscious cat is awakened by a mechanism that is
initiated by a radioactive decay.  In both cases an observer is permitted to check the statues of the cat at
any time during the experiment.  In all cases the nRules correctly and unambiguously predict the conscious
experience of the cat and the observer.
   
\end{abstract}

\section*{Introduction}
	Four rules called the \emph{nRules} are given in a previous paper \cite{RM1}.  These rules are said to govern
the process of stochastic choice and state reduction in an ontological model of a quantum mechanical system,
and describe how  conscious awareness of the observer changes (or not) during this process.  In the
present paper, these rules are applied to  Schr\"{o}dinger's cat.

	The first of the nRules refers to the probability current $J$ that flows into a state.  Current $J$ is
 the time rate of change of the square modulus.  

\vspace{.4cm}

\noindent
\textbf{nRule (1)}: \emph{For any subsystem of n components in a system having a total square modulus equal to
s, the probability per unit time of a stochastic choice of one of those components at time t is given by
$(\Sigma_nJ_n)/s$ where the net probability current $J_n$ going into the $n^{th}$ component at that time is
positive}.

\vspace{.4cm}

	The second rule specifies the conditions under which \emph{ready states} appear in solutions of
Schr\"{o}dinger's equation.  These are understood to be the \emph{basis states} of state reduction (i.e.,
collapse of the wave function in an individual case). Ready states are always underlined.

\vspace{.4cm}

\noindent
\textbf{nRule (2)}: \emph{If an interaction produces compact components that are discontinuous with the initial
component, then all of the new states that appear in these  components will be ready states}.

\noindent
[\textbf{note}: A \emph{compact component} is a single component that is not simultaneously continuous with
any other component.]

\leftskip=1cm
(see ref.\ 1 for elaboration on  ``discontinuous" and ``initial")

\leftskip=0cm
The third rule provides for the collapse of the wave.

\noindent
\textbf{nRule (3)}: \emph{If a component containing ready states is stochastically chosen, then all those
states will become realized, and all other components in the superposition will be immediately reduced to
zero.}

\noindent
[\textbf{note}: If a state in a compact component is not ready is called \emph{realized}. Realized states are
not underlined.]

\leftskip=1.9cm
(see ref. 1 for an elaboration on ``immediately")

\leftskip=0cm
\noindent
\textbf{nRule (4)}: \emph{If a component in a superposition is entangled with a ready state, then that
component can only receive probability current.}

\vspace{.4cm}

The purpose of the present paper is to apply these rules to two versions of Schr\"{o}dinger's cat experiment. 
Version I assumes that the cat is initially conscious and is made unconscious by  a mechanism
that is initiated by a radioactive decay.  In version II, the cat is initially unconscious and is made
conscious by an alarm clock that is initiated by a radioactive decay.

\section*{The Apparatus}
	We first look at the apparatus that is used in  Schr\"{o}dinger's cat experiment \emph{without} a cat or an
external observer being present.  It consists of a radioactive source and a detector (denoted by either $d_0$
or
$d_1$), where $d_0$ means that the detector has not yet captured the decay particle, and $d_1$ means that it
has.  The detector output will be connected to a mechanical device that carries out a certain task, such as a
hammer falling on a container that  releases an anesthetic gas.  This device is a function of time given
by $M(t)$, where $M(t_0)$ is its configuration (e.g., the position of the hammer) prior to its being
stochastically chosen.  The component $d_0M(t_0)$ indicates that the radioactive source had not yet triggered
the mechanism $M$.  The component $d_1M(t)$ indicates that the detector has captured a decay particle and the
mechanism has advanced to its position at time $t$.  Let $i_0$ be an indicator that tells us that $M$ has not
yet completed its task, and $i_1$ tell us that it has.  When $M = M(t_f)$ we will say the device has fully run
its course, so $d_1M(t_f)i_1$ means the source has decayed and the mechanical device has completed its task as
indicated by the time $t_f$ and the indicator $i_1$.  We also suppose that the source is exposed to the
detector for a time that is limited to the half-life of a single emission.  At that time a clock will shut off
the detector, so it will remain in the state $d_0$ if there has not yet been a particle capture. 	

\vspace{.4cm}

	The system is $\Phi(t_0) = d_0M(t_0)i_0$ at the beginning of the primary interaction at time $t_0$ when the
detector is first exposed to the radioactive source.  After that
\begin{equation}
\Phi(t \ge t_0) = d_0(t)M(t_0)i_0 + \underline{d}_1(t)M(t_0)i_0
\end{equation}
where the second component is zero at $t_0$ and increases in time. This component cannot evolve beyond $t_0$
because it is a ready state as required by nRule (2) and is therefore a \mbox{dead-end} because of nRule (4). 
We do not represent the source in this expression.  Probability  current flows from the first to the second
component, so it is possible that there will be a stochastic hit on the ready state $\underline{d}_1$ at time
$t_{sc}$ giving
\begin{displaymath}
\Phi(t = t_{sc} > t_0) = d_1(t)M(t_0)i_0
\end{displaymath}
where $d_1$ has been made a realized state by nRule (3).  As  the mechanical device  subsequently completes
its course at $t_f$, we will have
\begin{displaymath}
\Phi(t_f > t > t_{sc} > t_0) = d_1(t_{sc})M(t_0)i_0 \rightarrow d_1(t_{sc})M(t)i_0 \rightarrow
d_1(t_{sc})M(t_f)i_1 
\end{displaymath}
where the arrows denote a continuous evolution from $t_{sc}$ to $t_f$.  At and after $t_f$ the state of the
system is
\begin{equation}
\Phi(t \ge t_f > t_{sc} > t_0) = d_1(t_{sc})M(t_f)i_1
\end{equation}

If the half-life time $t_{1/2}$ of the source runs out before there is a stochastic hit, then eq.\ 1 will
become time independent and the second component will become a \emph{phantom}\footnote{A phantom no longer
serves any purpose and can be dropped from the equation.  This is like redefining the system at $t_{1/2}$. 
For further discussion, see ref.\ 1}.  The phantom can be dropped out of the equation and we are then left
with  
\begin{equation}
\Phi(t \ge t_{1/2} > t_0) = d_0(t_{1/2})M(t_0)i_0
\end{equation}

Fifty percent of the time the system will finish with $d_1(t_{sc})M(t_f)i_1$ in eq.\ 2, and the rest of the
time it will finish with $d_0(t_{1/2})M(t_0)i_0$ in eq.\ 3.

\section*{Add Observer}
	If an observer looks at the apparatus during the time before there has been a stochastic hit (eq.\ 1), then
prior to the observer's interaction we will have
\begin{equation}
\Phi(t \ge t_0) = \{d_0(t)M(t_0)i_0 + \underline{d}_1(t)M(t_0)i_0\}\otimes X
\end{equation}
where $X$ is the unknown state of the observer prior to interaction.  Let the observer look at the detector at
time $t_{look}$.  
\begin{eqnarray}
\Phi = d_0(t_{look})M(t_0)i_0\otimes X &\rightarrow& d_0(t_{look} + \pi)M(t_0)I_0B_0\\ 
+ \underline{d}_1(t_{look})M(t_0)i_0\otimes X &\rightarrow& \underline{d}_1(t_{look} + \pi)M(t_0)I_0B_0
\nonumber
\end{eqnarray}
where $B_0$ is the observer's brain when it is conscious of the indicator $I_0$.  The capital $I$ is an
extended indicator that includes the original device $i$ plus the low level physiological processes of the
observer.  The brain state $B$ only includes the part of the brain that is explicitly involved with the
conscious experience.  The continuous evolution (arrows) in eq.\ 5 carries $i$ into $I$ and $\otimes X$ into
$B_0$ in time $\pi$.  In this treatment arrows always indicate a continuous evolution, whereas a plus sign
indicates a discontinuous ``quantum jump".  

	The first row in eq.\ 5 is a \emph{single component} that evolves continuously  in response to the
physiological interaction.  The primary interaction is still active during this time and that gives rise to a
vertical current going from the first to the second row in eq.\ 5. The vertical evolution  is a discontinuous
jump in both cases.  The second row is therefore a continuum of components that are created parallel to the
first row at each moment of time.  So at  time $t_{look} + \pi$, vertical current flows only into the final
component in the second row of \mbox{eq.\ 5}.  Components prior to the last one no longer have current flowing
into them from above; and since there can be no horizontal current among these ready states, they become
phantom components as soon as they are created\footnote{Any component in the second row is compact.  It is
continuous with itself in time, but not with any \emph{other} component.  It is also a ready component arising
out of the primary interaction; and for this reason, it cannot undergo a continuous evolution on its own.  So
as soon as vertical current from the first row stops flowing into it, it becomes a phantom.  However, new
components arise that are `temporally' continuous with it.  The result is a continuum of ready states in the
second row that are phantoms except for the one that still receives vertical current.}.  The time of the
observation (i.e., when the physiological interaction is complete) is called $t_{ob} = t_{look} + \pi$, so
eq.\ 5 is
\begin{equation}
\Phi(t \ge  t_{ob} \ge t_0) = d_0(t)M(t_0)I_0B _0 + \underline{d}_1(t)M(t_0)I_0B_0
\end{equation}
where $\underline{d}_1(t)M(t_0)I_0B_0$ is zero at $t_{ob}$ and increases in time.  This is the same as eq.\ 4
with the observer on board.  

 With a stochastic hit on the second component in eq.\ 6 at time $t_{sc}$, we get
\begin{displaymath}
\Phi(t = t_{sc} > t_{ob} >  t_0) = d_1(t)M(t_0)I_0B_0 
\end{displaymath}
As the mechanical device runs its course from $t_{sc}$ to $t_f$ we will have
\begin{displaymath}
\Phi(t_f \ge t \ge t_{sc} > t_0) = d_1(t_{sc})M(t_0)I_0B_0 \rightarrow d_1(t_{sc})M(t)I_0B_0 \rightarrow
d_1(t_{sc})M(t_f)I_1B_1 
\end{displaymath}
where the arrows denote a continuous evolution from $t_{sc}$ to $t_f$.  The brain state $B_1$ observes the
indicator state $I_1$.  So prior to the stochastic hit, the observer in eq.\ 6 is conscious of $I_0$;  and
after the stochastic hit, the observer is first conscious of $I_0$ and then $I_1$. There are two different
brain states in this expression, but they occur at different times so there is no paradoxical ambiguity.   At
and after $t_f$ the state of the system is
\begin{equation}
\Phi(t \ge  t_f) = d_1(t_{sc})M(t_f)I_1B_1 
\end{equation}
which says that the observer has come on board and sees the final state of the indicator $I_1$.  

	If the clock runs out at $t_{1/2}$ before there has been a stochastic hit, then eq.\ 6 will become time
independent and the second component will become a phantom.  In that case  
\begin{equation}
\Phi(t \ge  t_{1/2} > t_{ob}) = d_0(t_{1/2})M(t_0)I_0B _0 
\end{equation}
Fifty percent of the time the system will finish with $d_1(t_f)M(t_f)I_1B_1$ in eq.\ 7, and the rest of the
time it will finish with $d_0(t_{1/2})M(t_0)I_0B _0$ in eq.\ 8.

If there should be a stochastic hit on eq.\ 5 in the middle of the physiological interaction (i.e., the
continuous horizontal development in that equation), then the corresponding ready state in the second row will
be chosen, and a subsequent classical/continuous evolution will carry the system all the way to eq.\ 7.

\section*{Version 1 with no Outside Observer}
	We now replace the indicator in eq.\ 1 with cat brain states, the first of which is the conscious state $C$
shown in eq.\ 9 below.  This says that the cat is initially conscious of the mechanical device in its
 state $M(t_0)$.  The device is understood to render the cat unconscious when it reaches $M(t_f)$.  In this
case we require that all lower physiological operations of the cat's brain are included in mechanical device. 
Before a stochastic choice occurs, the system is given by
\begin{equation}
\Phi(t \ge  t_0) = d_0(t)M(t_0)C + \underline{d}_1(t)M(t_0)C
\end{equation}
where the second component is zero at $t_0$ and increases in time, and where the cat is entangled with the
mechanical device from the beginning.  The ready detector state $\underline{d}_1$ represents a capture of the
radioactive decay, the source of which is not shown.  If there is a stochastic hit on the second component at
time $t_{sc}$, eq.\ 9 becomes
\begin{displaymath}
\Phi(t_f \ge t \ge t_{sc} > t_0) = d_1(t_{sc})M(t_{sc})C \rightarrow d_1(t_{sc})M(t)C \rightarrow
d_1(t_{sc})M(t_f)U 
\end{displaymath}
where the arrows represent a continuous classical evolution from $M(t_{sc})$ to $M(t_f)$ and from the conscious
state $C$ to the unconscious state $U$. Again, there are two different brain states in this expression, but
they occur at different times so there is no paradoxical ambiguity of the kind generally associated with
Schr\"{o}dinger's cat.

If there is no stochastic hit on eq.\ 9 by the time the interaction is cut off at $t_{1/2}$, then the second
component will become a phantom, leaving
\begin{displaymath}
\Phi(t \ge t_{1/2} > t_0) = d_0(t_{1/2})M(t_0)C
\end{displaymath}
So the cat will have escaped unconsciousness 50\% of the time

\section*{Version 1 with Outside Observer}
We begin with eq.\ 9 except that there is now an outside observer $X$ waiting in the wings 
\begin{displaymath}
\Phi(t \ge  t_0) = \{d_0(t)M(t_0)C + \underline{d}_1(t)M(t_0)C\}\otimes X
\end{displaymath}
Let the observer look at the cat at time $t_{look}$. 
\begin{eqnarray}
\Phi = d_0(t_{look})M(t_0)C\otimes X &\rightarrow& d_0(t_{ob})M(t_0)CB_C\nonumber\\ 
+ \underline{d}_1(t_{look})M(t_0)C\otimes X &\rightarrow& \underline{d}_1(t_{ob})M(t_0)CB_C\nonumber
\end{eqnarray}
where $B_C$ is the observer's brain when it is conscious of the  cat in its conscious state $C$. The
mechanical device is now expanded to include the low level physiology of the outside observer.  As in eq.\ 5,
the second row is a continuum of ready components, only the last of which survives at $t_{ob}$ to give 
\begin{equation}
\Phi(t \ge  t_{ob} >  t_0) = d_0(t)M(t_0)CB_C + \underline{d}_1(t)M(t_0)CB_C
\end{equation}
where the second component is zero at $t_{ob}$ and increases in time.  If there is a stochastic hit on the
second component of this equation, we will get
\begin{displaymath}
\Phi(t_f \ge t \ge t_{sc} >  t_{ob} > t_0) =   
\end{displaymath}
\begin{displaymath}
 =d_1(t_{sc})M(t_{sc})CB_C \rightarrow d_1(t_{sc})M(t)CB_C \rightarrow d_1(t_{sc})M(t_f)UB_U
\end{displaymath}
where $B_U$ is the observer's  brain state when it is conscious of the unconscious cat in the ready state
$U$.  This equation says that at time $t_f$ the cat is made unconscious and is observed in that state by
the outside observer.  In this case both the cat and the observer have different brain states in the same
expression, but they both occur at different times so the result is not paradoxical. 

	If there is no stochastic hit on eq.\ 10 by the cut-off time $t_{1/2}$, then the second component will become
a phantom, leaving
\begin{displaymath}
\Phi(t  \ge t_{1/2} > t_{ob} > t_0) = d_1(t_{1/2})M(t_0)CB_C 
\end{displaymath}
indicating that the cat has escaped unconsciousness as observed by the outside observer.

\section*{Version II with no Outside Observer}
	In the second version of the Schr\"{o}dinger cat experiment, the cat is initially unconscious and is awakened
by an alarm that is set off by the capture of a radioactive decay.  The mechanical device $M(t)$ is now an
alarm clock.  As before, it will go off when the device reaches $M(t_f)$, which happens only 50\% of the
time.  
\begin{equation}
\Phi(t \ge t_0) = d_0(t)M(t_0)U + \underline{d}_1(t)M(t_0)U
\end{equation}
If there is a stochastic hit on the second component, then
\begin{displaymath}
\Phi(t_f \ge t \ge  t_{sc} > t_0) = d_1(t_{sc})M(t_{sc})U \rightarrow d_1(t_{sc})M(t)U \rightarrow
d_1(t_{sc})M(t_f)C
\end{displaymath}
so
\begin{displaymath}
\Phi(t \ge t_f >  t_{sc}) = d_1(t_{sc})M(t_f)C
\end{displaymath}

	If there is no stochastic hit by the time $t_{1/2}$, then the second component in eq.\ 11 is a phantom and we
get just
\begin{equation}
\Phi(t \ge t_{1/2} >  t_0) = d_0(t_{1/2})M(t_0)U 
\end{equation}

\section*{Version II with Outside Observer}
	Starting again with eq.\ 11 with an outside observer standing by
\begin{displaymath}
\Phi(t \ge   t_0) = \{d_0(t)M(t_0)U + \underline{d}_1(t)M(t_0)U\}\otimes X
\end{displaymath}
The observer looks at time $t_{look}$
\begin{eqnarray}
\Phi = d_0(t_{look})M(t_0)U\otimes X &\rightarrow& d_0(t_{ob})M(t_0)UB_U\nonumber\\ 
+ \underline{d}_1(t_{look})M(t_0)U\otimes X &\rightarrow& \underline{d}_1(t_{ob})M(t_0)UB_U\nonumber
\end{eqnarray}
In the second row, only the last component  survives to give
\begin{equation}
\Phi(t \ge t_{ob} >  t_0) = d_0(t)M(t_0)UB_U + \underline{d}_1(t)M(t_0)UB_U
\end{equation}
which puts the observer on board with the unconscious cat.  The second component is zero at $t_{ob}$ and
increases in time.  If there is then a stochastic hit at time $t_{sc}$
\begin{displaymath}
\Phi(t = t_{sc} > t_{ob} >  t_0) = d_1(t_{sc})M(t_0)UB_U
\end{displaymath}
and subsequently
\begin{displaymath}
\Phi(t_f > t \ge t_{sc}) = d_1(t_{sc})M(t_0)UB_U \rightarrow d_1(t_{sc})M(t)UB_U \rightarrow
d_1(t_{sc})M(t_f)CB_C
\end{displaymath}
concluding in
\begin{displaymath}
\Phi(t \ge t_f ) = d_1(t_{sc})M(t_f)CB_C
\end{displaymath}
If there is no stochastic hit, then the second component in eq.\ 13 is a phantom and we are left with
\begin{displaymath}
\Phi(t \ge t_{1/2} > t_{ob} > t_0 ) = d_0(t_{1/2})M(t_0)UB_U
\end{displaymath}

\section*{Version II with a Natural Wake-Up}
Even if the alarm does not go off, the cat will wake up naturally by virtue of its own internal alarm clock. 
The internal alarm can be represented by a classical mechanism $N(t)$ that operates at the same time as the
external alarm $M(t)$.  The interaction  runs parallel to eq.\ 11 and is given by
\begin{equation}
\Phi(t \ge  t_0 ) = N(t_0)U \rightarrow N(t)U  \rightarrow N(t_f)C
\end{equation}
Taking the product of eqs.\ 11 and 14 at $t_0$ gives
\begin{displaymath}
\Phi(t =  t_0 ) = d_0(t)M(t_0)N(t_0)U 
\end{displaymath}
after which 
\begin{displaymath}
\Phi(t \ge t_0 ) = [d_0(t)M(t_0)U + \underline{d}_1(t)M(t_0)U][N(t_0)U \rightarrow  N(t)U \rightarrow N(t_f)C]
\end{displaymath}
where the cross product suggests a conflict between $C$ and $U$ states.  To resolve this, we follow two
possible scenarios.  The first assumes that the external stochastic choice and decay  occurs
before the internal decay, and the second assumes that the internal decay occurs before the
external stochastic choice and decay.  The first of these gives
\begin{displaymath}
\Phi(t \ge t_0 ) = d_0(t)M(t_0)[N(t_0) \rightarrow N(t)]U + \underline{d}_1(t)M(t_0)[N(t_0) \rightarrow N(t)]U
\end{displaymath}
Following a stochastic hit we have
\begin{displaymath}
\Phi(t >t_{ff} > t_f > t_{sc} > t_0 ) =
\end{displaymath}
\begin{displaymath}
=[d_1(t_{sc})M(t_0)U \rightarrow d_1(t_{sc})M(t)U \rightarrow
d_1(t_{sc})M(t_f)C][N(t_0)
\rightarrow N(t)]
\end{displaymath}
resulting in
\begin{eqnarray}
\Phi(t >t_{ff} > t_f > t_{sc} > t_0 ) &=&  d_1(t_{sc})M(t_f)[N(t_0) \rightarrow N(t) \rightarrow N(t_{ff})]C
\nonumber\\ &=& d_1(t_{sc})M(t_f)N(t_{ff})C
\end{eqnarray}
where $t_{ff}$ is the final time of the  internal continuous development.

The second scenario is
\begin{displaymath}
\Phi(t \ge t_0 ) = [N(t_0)U \rightarrow N(t)U \rightarrow N(t_{ff})C][d_0(t)M(t_0) +
\underline{d}_1(t)M(t_0)]
\end{displaymath}
resulting in
\begin{displaymath}
\Phi(t \ge t_{ff} > t_0) = N(t_{ff})[d_0(t)M(t_0) + \underline{d}_1(t)M(t_0)]C
\end{displaymath}
followed by a stochastic hit, giving 
\begin{displaymath}
\Phi(t_f > t > t_{sc} >  t_{ff} > t_0 ) = N(t_{ff})d_1(t_{sc})[M(t_{sc}) \rightarrow M(t) \rightarrow M(t_f)]C
\end{displaymath}
resulting in
\begin{equation}
\Phi(t > t_f > t_{sc} >  t_{ff} > t_0 ) = d_1(t_{sc})M(t_f)N(t_{ff})C 
\end{equation}
Both of these scenarios lead to the same conscious state in eqs.\ 15 and 16.

\end{document}